\documentclass[preprint,superscriptaddress,showpacs,preprintnumbers,amsmath,amssymb]{revtex4}

\usepackage{graphicx}
\usepackage{dcolumn}
\usepackage{bm}
\usepackage{rotating}

\begin{document}
\title{Counting Solutions for the $N$-queens and Latin Square Problems by Efficient Monte Carlo Simulations}

\author{Cheng Zhang}
\affiliation{Department of Bioengineering, Rice
University, Houston, Texas 77005, USA}

\pacs{02.10.Ox, 05.10.Ln, 75.40.Mg}

\author{Jianpeng Ma}
\email{jpma@bcm.tmc.edu} \affiliation{Department of Bioengineering,
Rice University, Houston, Texas 77005, USA}

\affiliation{ Verna and Marrs McLean Department of Biochemistry and
Molecular Biology, Baylor College of Medicine, One Baylor Plaza,
BCM-125, Houston, Texas 77030, USA }

\begin{abstract}
We apply Monte Carlo simulations to count the numbers of solutions
of two well-known combinatorial problems: the $N$-queens problem and
Latin square problem.  The original system is first converted to a
general thermodynamic system, from which the number of solutions of
the original system is obtained by using the method of computing the
partition function.  Collective moves are used to further accelerate
sampling: swap moves are used in the $N$-queens problem and a
cluster algorithm is developed for the Latin squares.  The method
can handle systems of $10^4$ degrees of freedom with more than
$10^{10000}$ solutions. We also observe a distinct finite size
effect of the Latin square system: its heat capacity gradually
develops a second maximum as the size increases.

\end{abstract}

\maketitle

\begin{figure}[h]
    \begin{minipage}{ 0.45 \linewidth}
    (a)
        \begin{center}
            \includegraphics[width=  \linewidth]{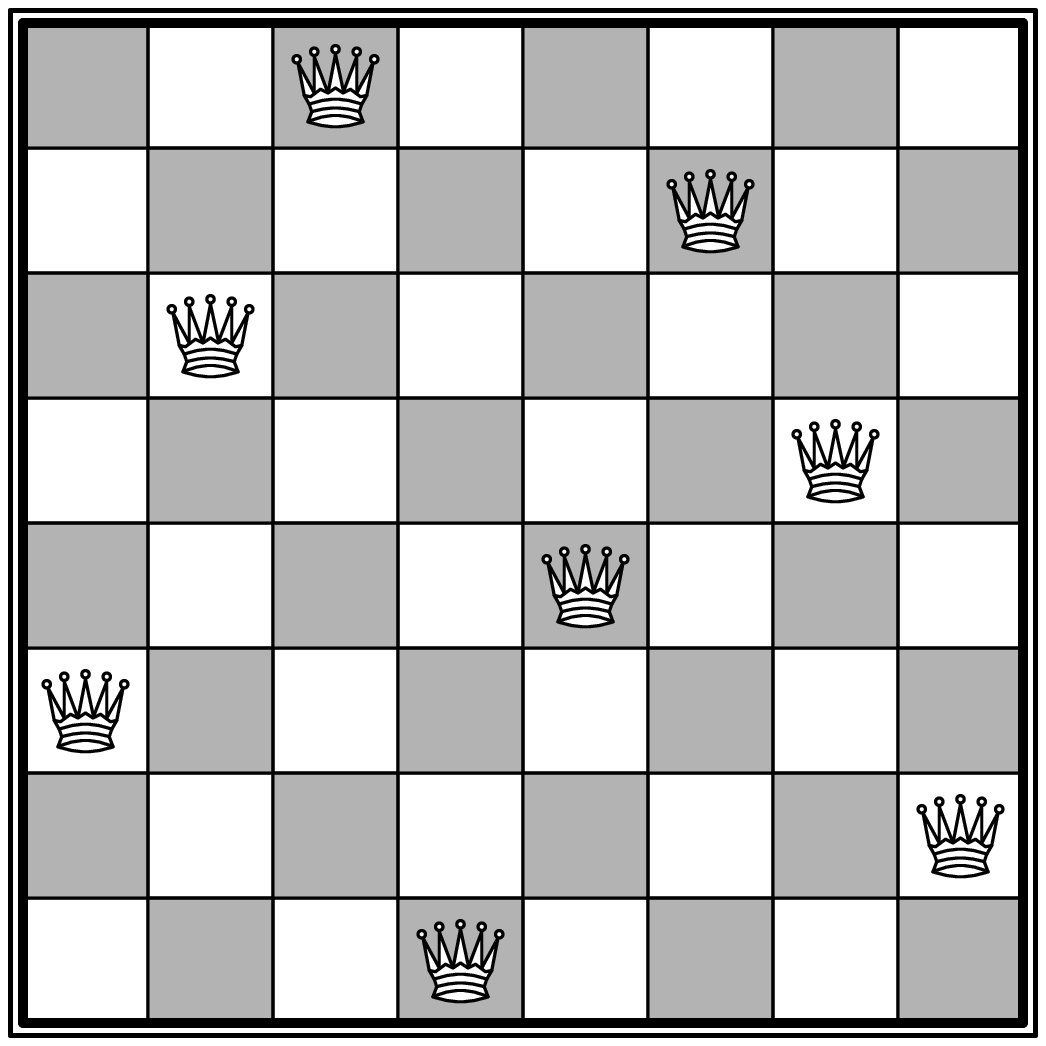}
        \end{center}
    \end{minipage}
    \begin{minipage}{ 0.45 \linewidth}
    (b)
        \begin{center}
            \includegraphics[width=  \linewidth]{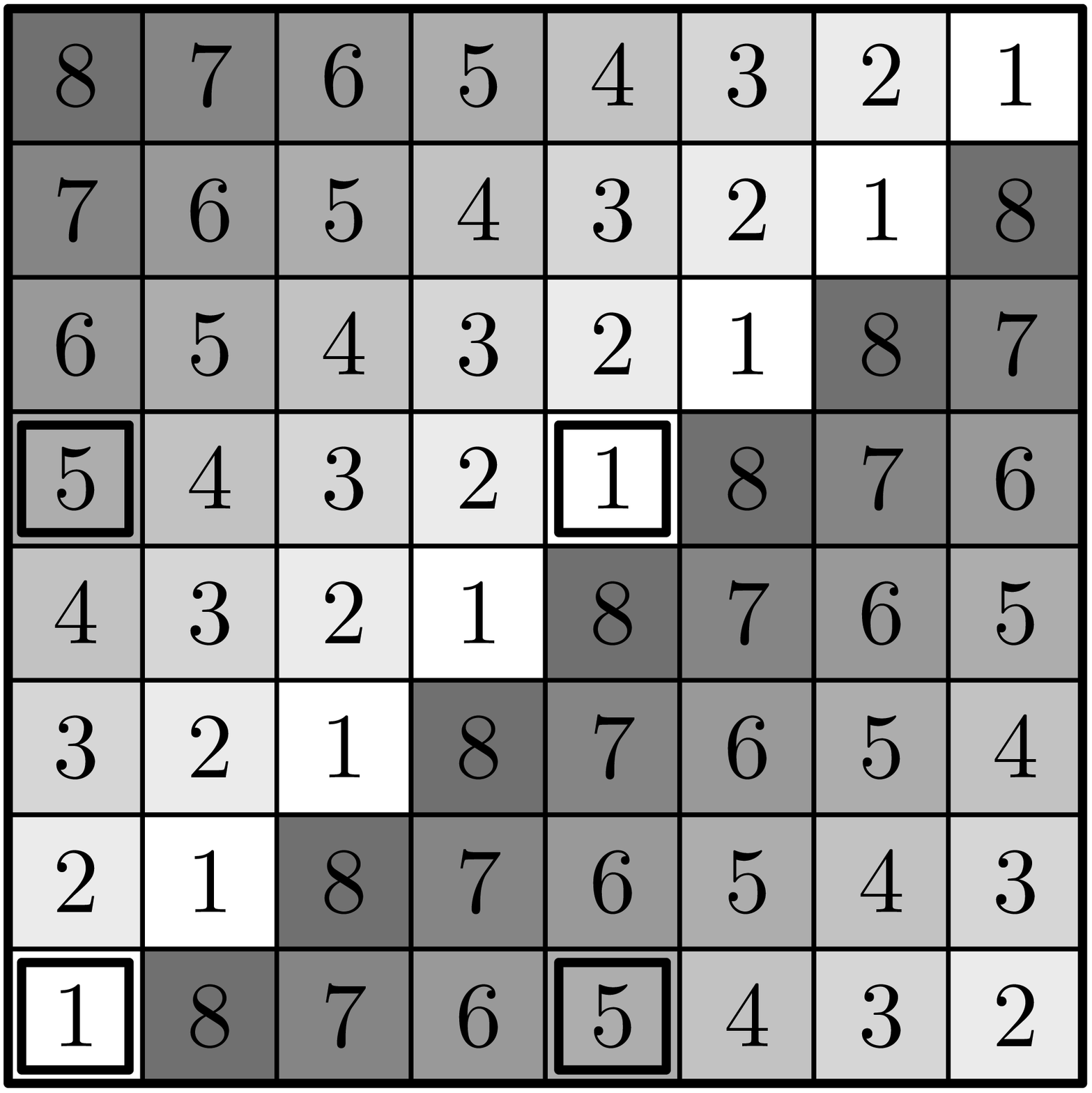}
        \end{center}
    \end{minipage}
  \caption{\label{fig:queen} (a) In the $N$-queens problem, a solution is a way of placing
  $N$ (here $N=8$) queens on an $N\times N$ chessboard such that no two queens attack
each other horizontally, vertically or diagonally.
  (b) The Latin square problem requires one to use $L$ different symbols
  (in this case $L=8$ and the symbols are $1,2, \ldots, L$)
  to fill an $L\times L$ table such that each symbol only occurs once
  in  any row/column.
  An example of a cluster generated by the cluster
  algorithm (see text) is shown by the four marked cells. After it is generated,
  the symbols `1' and `5' within the cluster are exchanged.}
\end{figure}

Counting solutions of constraint-satisfaction problems is a
fundamental subject in basic science and engineering. Specifically,
one aims at calculating the number of ways for a system to satisfy a
set of constraints simultaneously.  For example, in the $N$-queens
problem, the constraints are to avoid $N$ queens on an $N\times N$
chessboard attacking one another, see Fig. \ref{fig:queen}(a).  In
the Latin square problem, one looks for ways of filling an $L\times
L$ table using $L$ different symbols such that in every row or
column, each symbol only occurs once, see Fig.~\ref{fig:queen}(b).

As standard benchmark tests, many heuristic and combinatorial
methods are developed to search for one or a few of their solutions,
e.g., the min conflicts algorithm \cite{b.minconflict}, dynamic
programming \cite{b.dp} and iterated map method \cite{b.itermap}.
However, to count all solutions is a more challenging task.  The
traditional approaches by a complete enumeration in general can only
handle systems of a relatively small size because the number of
solutions grows exponentially with the system size.  To date, the
largest system ($N=25$) of the $N$-queens problem contains about
$2.21\times 10^{15}$ solutions according to a recent enumeration
\cite{b.nqueenexact}. For the Latin square problem, the largest
exactly-solved system $L=11$ has about $7.77\times10^{47}$ solutions
\cite{b.latinexact}.

An alternative approach is to calculate the ratio between the number
of solutions of the original problem and that of a simplified
problem.  If we know the exact number of solutions of the simplified
problem, then the number of solutions of the original problem can be
deduced.

To connect the original problem (denoted as $O$) with the simpler
problem (denoted as $S$), we carefully choose the problem $S$ to be
a generalized version of the problem $O$ such that every solution of
the problem $O$ is a solution of the problem $S$.  Here, the simpler
problem $S$ typically has fewer constraints, and hence more
(easier-to-find) solutions.  We then perform a Monte Carlo
simulation in the configurational space spanned by all solutions of
the problem $S$ to compute the ratio of solutions of $O$  and $S$.
A convenient way to recognize a solution of the problem $O$ is to
use an energy function $E$ that is nonnegative everywhere and is
zero if and only if the configuration is a solution of the problem
$O$.

Since the numbers of solutions of $O$  and  $S$ usually differ by
many orders of magnitudes as the system size increases, the ratio of
the two becomes too small to be computed directly. Therefore we need
a set of intermediate problems $\{S_i\}$, each of which is
associated with a reciprocal temperature  $\beta_i$.  The $\beta_i$
weights each configuration according to its energy $E$ as
$\exp(-\beta _i E)$. The weighted sum of solutions using $\beta_i$
is the partition function $Z_i=\sum \exp(-\beta_i E)$. Note, the
partition function has an interpretation of the number of solutions
in two extreme cases: the number of solutions of the problem $S$
corresponds to the partition function at $\beta=0$, and that of the
problem $O$ is the partition function at $\beta\rightarrow \infty$,
where only zero-energy configurations can survive. Several Monte
Carlo methods were previously used to infer the partition function
\cite{b.mc}. However, these methods failed to be applied to large
systems.


To handle large systems, we use a Monte Carlo method that directly
computes the partition function \cite{b.Z}, where we simultaneously
sample the system at multiple temperatures by means of transitions
between the temperatures.
In addition to configurational space
sampling under a fixed temperature $\beta_i$, e.g. the Metropolis
algorithm \cite{b.metropolis}, temperature transitions are randomly
proposed from the current value $\beta_i$ to another one $\beta_j$,
and accepted with a probability
$\mathrm{Acc}_\beta=\min\{1,\exp[-(\beta_j-\beta_i) E+\ln \tilde
Z_i-\ln \tilde Z_j]\}$.
Here, $E$ is current energy, $\tilde Z_i$ and $\tilde Z_j$ are the
estimated values of the partition function at $\beta_i$ and
$\beta_j$, respectively. %
The  $\ln \tilde Z_i$'s are then dynamically converged to the actual
values $\ln Z_i$'s through a recursive updating until an accuracy
$|\ln \tilde Z_i-\ln Z_i|<0.10$ is reached \cite{b.Z}. This accuracy
guarantees a correct order of magnitude of the $\tilde Z_i$ (which
is $\log_{10} \tilde Z_i=\ln \tilde Z_i/\ln{10}$). To obtain a more
accurate partition function, we perform an additional run of
simulation with all  $\tilde Z_i$'s fixed at their final values.
Practically, the final run is always much longer than all the
previous updating stages; thus the cost of the updating is
negligible. The statistics accumulated from the final run is used to
further refine the partition function through the multiple histogram
method \cite{b.multihist}.

\begin{figure}[h]
  \begin{minipage}{ \linewidth}
    \begin{center}
        \includegraphics[angle=-90,width=  \linewidth]{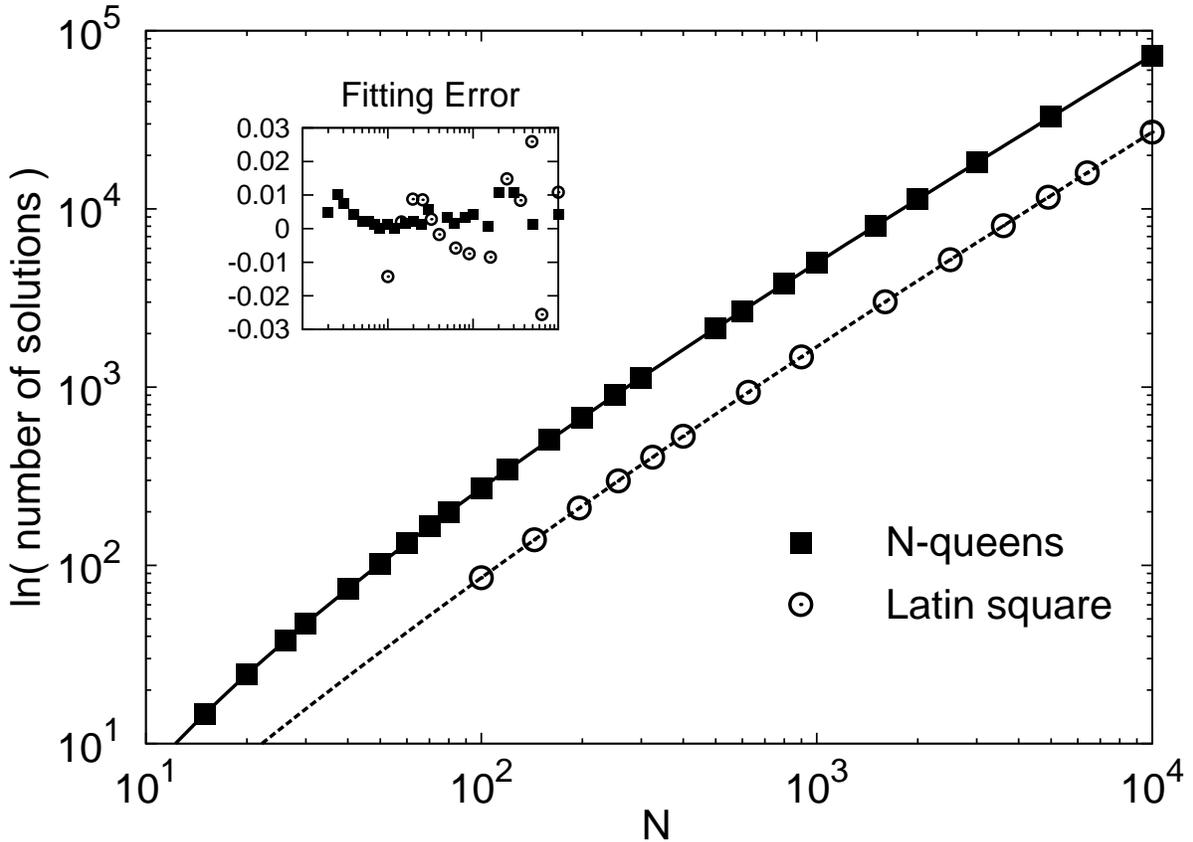}
    \end{center}
  \end{minipage}%
  \caption{\label{fig:nqueens}
The numbers of solutions of the $N$-queens problem $Q_N$ and that of
the Latin square problem $S_L$ versus the system size $N$ (for a
Latin square $N=L\times L$). There is a simple linear relation
between $\ln (N!/Q_N)$ and $N$ while a fitting formula for $S_L$ is
more complicated (see text). The inset shows the error of fitting
the formulas to the numerical results. }
\end{figure}

For the $N$-queens problem, see Fig. \ref{fig:queen}(a), the
$N$-rooks problem can serve as the problem $S$, where queens
function as rooks such that they can attack each other only
horizontally and vertically, but not diagonally. The problem $S$ is
a trivial one: each of its solutions corresponds to a permutation of
the $N$ column indices because the row constraints are satisfied by
placing only one rook on each row while the column constraints are
satisfied by placing rooks from different rows at different columns.
Hence, there are totally $N!$ solutions for the $N$-rooks problem.

We now specify the energy function that connects the simple problem
with the original one.  If the diagonal $d$ has $C_d$ resident
queens, the energy of that diagonal $E_d = \max\{C_d-1, 0\}$. The
energy of the whole system is a sum of the energy of all diagonals.
A zero-energy configuration guarantees that no diagonal has more
than one queen, and therefore is a solution of the $N$-queens
problem.

We used the swap move introduced by Sosic and Gu \cite{b.swapmove}
to sample the configurational space.  In each Monte Carlo step, we
randomly choose two rows and try to swap the column indices of the
queens there.  Note, after a swap the horizontal and vertical
constraints are still satisfied.  Thus these swaps can be used to
perform sampling on the configurational space of the problem $S$.

The number of solutions for systems of several typical sizes are
shown in Table \ref{tab.nqueens}.  For the largest exactly-solved
system to date $N=25$ \cite{b.nqueenexact}, the relative error is
only $5\times 10^{-5}$.  The results on small systems serve as a
check of our method. Currently, there is a dispute about the number
of solution for $N=24$. An alternative calculation \cite{b.n24alt}
gives 226732487925864 solutions instead of the value 227514171973736
used in Table~\ref{tab.nqueens}.  Our long-time simulation result
$2.2751\times 10^{14}$ clearly supports the latter result.  More
importantly, our method can be used on much larger systems, to which
one cannot apply traditional counting algorithms due to
astronomically large numbers of solutions.  In the largest system,
there are about $1.33\times 10^{31560}$ solutions for $N=10000$ (in
which case we used 82 temperatures from $\beta=9.2$ to $\beta=0$).
The results on large systems are shown in Fig.~\ref{fig:nqueens}.
Our linear fitting result shows that for large systems $N > 100$,
the number of solutions $Q_N$ satisfies $\ln(N!/Q_N) \approx
0.944001 N - 0.937$; the maximal fitting error is less than 0.02 in
this range.

\begin{table}
\caption{The numbers of solutions $Q_N$ of the $N$-queens problems.
The simulation cost are measured by sweeps (numbers of Monte Carlo
steps per queen). The first six significant digits of the exact
results \cite{b.nqueenexact} are displayed in the last column for
comparison.} \label{tab.nqueens}
\begin{center}
\begin{tabular}{r|c|l|l}
$N$  & sweeps & $Q_N$  & exact value\\
\hline
21     & $4\times10^{10}$    & $3.1468\times10^{11}$ & $3.14666\times10^{11}$\\
22     & $5\times10^{10}$    & $2.6910\times10^{12}$ & $2.69101\times10^{12}$\\
23     & $4\times10^{10}$    & $2.4234\times10^{13}$ & $2.42339\times10^{13}$ \\ 
24     & $1\times10^{11}$    & $2.2751\times10^{14}$ & $2.27514\times10^{14}$ \\
25     & $1\times10^{11}$    & $2.2080\times10^{15}$ & $2.20789\times10^{15}$ \\ 
26     & $1\times10^{11}$    & $2.2320\times10^{16}$ &  \\
27     & $5\times10^{10}$    & $2.3489\times10^{17}$ &  \\
28     & $5\times10^{10}$    & $2.5645\times10^{18}$ &  \\
29     & $5\times10^{10}$    & $2.8899\times10^{19}$ &  \\
30     & $5\times10^{10}$    & $3.3731\times10^{20}$  &  \\
40     & $2\times10^{10}$    & $8.273\times10^{31}$  &  \\
50     & $2\times10^{10}$    & $2.456\times10^{44}$  &  \\
100    & $1\times10^{10}$    & $2.392\times10^{117}$ &  \\
200    & $1\times10^{10}$    & $2.041\times10^{293}$ &  \\
500    & $1\times10^{10}$    & $3.219\times10^{929}$ &  \\
1000   & $5\times10^{9}$     & $1.094\times10^{2158}$ & \\
2000   & $2\times10^{9}$     & $9.44\times10^{4915}$ & \\
5000   & $1\times10^{9}$     & $1.46\times10^{14276}$ & \\
10000  & $1\times10^{9}$     & $1.33\times10^{31560}$ &
\end{tabular}
\end{center}
\end{table}

Next, we turn to the Latin square problem.  For convenience we
choose $1,2,\ldots,L$  as the $L$ different symbols to fill the
$L\times L$ table. To construct a problem $S$, we remove the
constraints for columns, i.e., we no longer require each symbol to
occur once in a column, while retaining the constraints for rows.
Thus different rows act independently.  The constraints for symbols
within a row being mutually different imply that each row
configuration is a permutation of the $L$ symbols.  Thus there are
$L!$ different arrangements for each individual row, and $(L!)^L$
arrangements for the whole system (the problem $S$).

The energy function is the following.  A symbol that is shared by
two different rows on the same column contributes +1 to the total
energy, i.e., $E=\sum_{i<j;~ k}\delta(s_{ik}, s_{jk})$. Here,
$s_{ij}$ is the symbol at the $i$th row and $j$th column;
$\delta(a,b)$ is +1 if the two symbols $a$ and $b$ are the same,
zero otherwise; the two indices $i$ and $j$ enumerate over every
pair of different rows, $k$ every column. A Metropolis way to sample
the system is to randomly choose two columns on a row, and to try to
swap their symbols. Similar to the previous case, the swaps preserve
the constraints for rows, thus are qualified as a sampler of the
configurational space.

However, at a low temperature, the swap becomes inefficient due to
frequent rejections. For example, at the lowest temperature
$\beta=8.4$  we used for the $100\times100$ system, the average
probability of accepting a swap is less than 0.01\%. To overcome the
difficulty, we developed a rejection-free cluster algorithm for this
system and used it to generate configurational changes.  The cluster
algorithm is of the same spirit of its counterpart on the Ising
model \cite{b.cluster}.  It exploits the symmetry between any two
symbols $a$ and $b$, e.g., the system energy is unchanged if we
exchange the two symbols in a suitable collection of rows (or a
cluster).

A cluster is generated as the following.  We first randomly choose
two symbols $a$ and $b$ as well as a row index $i$, and add this row
index $i$ into the cluster as a ``seed''.  We now scan the row $i$
and pick up the column $j$ where the symbol $s_{ij}$  is $a$, and
search in other rows $i'$ for the symbol $b$ at the same column $j$,
i.e., $s_{i'j}=b$ . For each row $i'$ found, we use a probability
$P_{\mathrm{add}} = 1-\exp(-\beta)$ to add it into the cluster.
Similarly, we pick up the column $k$ where $s_{ik}=b$, and add every
other row $i''$ where $s_{i''k}=a$ to the cluster using the same
probability. This process is repeated until every row in the cluster
is considered. An example is shown in Fig. \ref{fig:queen}(b), where
$a=1$ and $b=5$, and the bottom row is the seed.  Once the cluster
is formed, we exchange the symbols $a$ and $b$ within.

The number of solutions of the Latin square problem is listed in
Table \ref{tab.latin}.  We used the Metropolis moves for small
systems, but cluster moves for large systems at low temperatures. In
this way we could access large systems, as shown in Fig.
\ref{fig:nqueens}. The size of the largest system is 100 by 100, in
which there are over $10^{11710}$  solutions.  In this system, we
used 85 temperatures from $\beta=8.4$ to $\beta=0$.  We attempted to
fit the number of solutions $S_L$ to the formula
$\ln(L!^L/S_L)\approx
L^2(0.99649+42.9721/L-35.8277/L^2)/(1+49.6514/L+152.80/L^2)$; the
maximal fitting error is 0.03.

\begin{table}
\caption{The numbers of solutions $S_L$  of the $L \times L$ Latin
square problems. One sweep is defined as the numbers of Monte Carlo
steps per site. The exact results \cite{b.latinexact} are displayed
to the first five significant digits. We used the cluster algorithm
for the last two systems.} \label{tab.latin}
\begin{center}
\begin{tabular}{c|l|l|l}
size  & sweeps & $S_L$  & exact value\\
\hline
$10\times10$    & $1\times10^{10}$  & $9.988\times10^{36}$ &  $9.9824\times10^{36}$ \\ 
$11\times11$    & $1\times10^{10}$  & $7.773\times10^{47}$ &  $7.7697\times10^{47}$ \\
$12\times12$    & $1\times10^{10}$  & $3.102\times10^{60}$ &   \\
$13\times13$    & $1\times10^{10}$  & $7.500\times10^{74}$ &   \\
$14\times14$    & $1\times10^{10}$  & $1.266\times10^{91}$ &   \\
$15\times15$    & $1\times10^{10}$  & $1.728\times10^{109}$ &  \\
$16\times16$    & $1\times10^{10}$  & $2.161\times10^{129}$ &   \\
$17\times17$    & $1\times10^{10}$  & $2.804\times10^{151}$ &   \\
$18\times18$    & $1\times10^{10}$  & $4.256\times10^{175}$ &   \\
$19\times19$    & $1\times10^{10}$  & $8.354\times10^{201}$ &   \\
$20\times20$    & $1\times10^{10}$  & $2.365\times10^{230}$ &   \\
\hline
$50\times50$    & $1\times10^{8}$   & $5.67\times10^{2250}$ &   \\
$100\times100$  & $1\times10^{7}$   & $1.55\times10^{11710}$ &   \\
\end{tabular}
\end{center}
\end{table}

\begin{figure}[h]
  \begin{minipage}{ \linewidth}
    \begin{center}
        \includegraphics[angle=-90,width=  \linewidth]{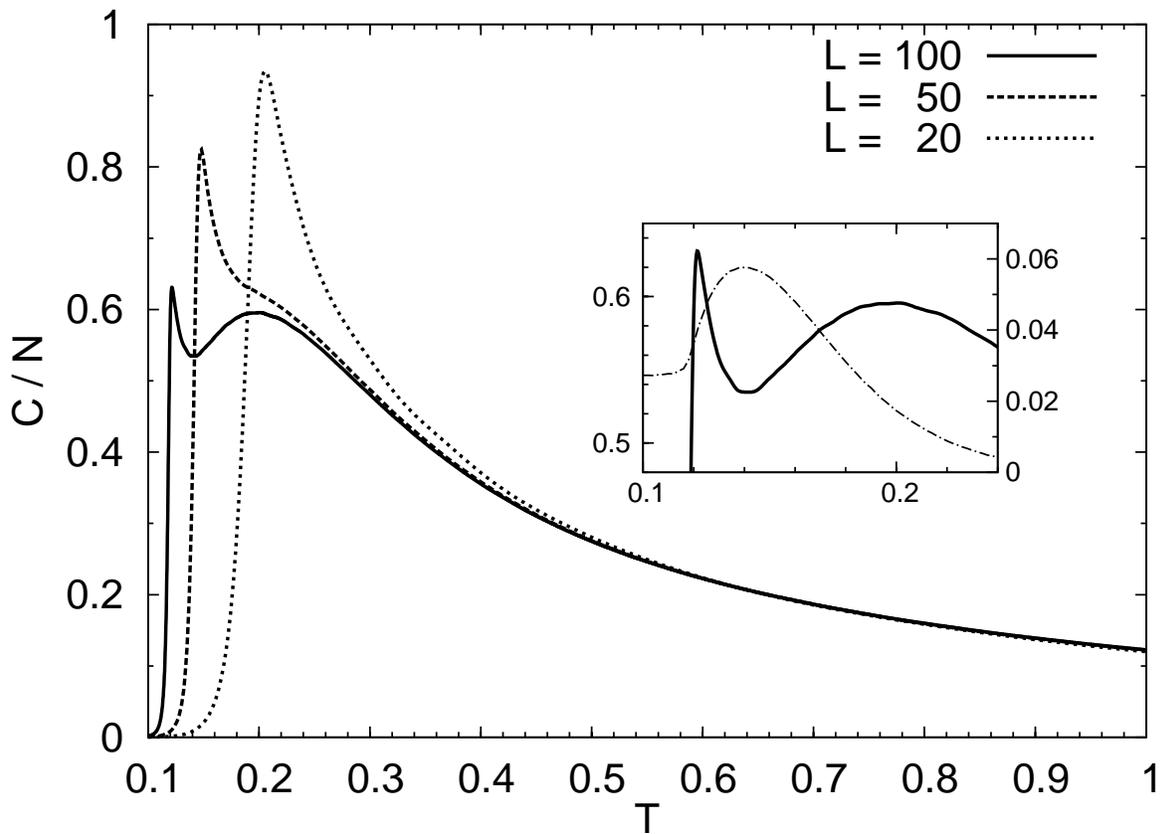}
    \end{center}
  \end{minipage}%
  \caption{\label{fig:cv}
Heat capacity $C$ per site of Latin squares versus temperature $T$.
The heat capacity develops two peaks as one increases the system
size. The inset shows that the valley between the two maxima of the
heat capacity for the $100\times 100$ system (the solid line, the
left axis) corresponds to where the fraction of percolated clusters
(the dash dot line, the right axis) reaches the maximum.}
\end{figure}

The heat capacity $C$ of the system shows an interesting finite-size
effect.  As the system size increases, the system develops two
separate maxima, see Fig. \ref{fig:cv}.
The anomaly of the heat capacity is a result of many frustrated low
energy states.  A similar phenomenon was experimentally observed in
a two-dimensional antiferromagnetic system \cite{b.antiferro}.
Besides, the valley between the two maxima coincides with the
location where
the system has the maximal fraction of percolated clusters. In the
cluster algorithm, a cluster is defined as percolated if it includes
all rows.
%
As shown in the inset of Fig. \ref{fig:cv}, for the $100 \times 100$
Latin square, the maximum fraction 0.06 occurs at $T_h \approx
0.14$, where the heat capacity hits its local minimum.
We now give a qualitative explanation for why the highest
percolation fraction occurs at a finite temperature $T_h$ rather
than $T=0$. At a very low temperature, each column has at most two
cells with the two symbols under concern ($a$ and $b$). As the
temperature is increased to $T_h$, a column is allowed to have more
of these cells. Meanwhile $P_\mathrm{add}$ is not changed
significantly from 1.0 (in the above example,
$P_\mathrm{add}\approx0.9992$ at $T_h$). Thus clusters are more
readily spread over rows than at $T=0$. However, a further increase
of the temperature decreases $P_\mathrm{add}$ and suppresses the
growth of clusters.

In summary, we demonstrate an efficient method to count the number
of solutions for the $N$-queens problem and Latin square problem.
The original problem $O$ is generalized to a less-constrained
problem $S$ and its partition function is calculated. We achieved a
high sampling efficiency by using collective Monte Carlo moves: in
the $N$-queens problem, the column indices of the queens are swapped
rather than altered individually; similarly, in the Latin square
problem, symbols within a row are always exchanged (the cluster move
is even more collective because we also attempt to exchange symbols
in different rows). These collective moves not only improve the
sampling efficiency at low temperatures, but also reduce the
sampling space by making the problem $S$ as close to the problem $O$
as possible. In the $N$-queens problem, the use of the swap move
reduces the sampling space of the problem $S$ from $N^N$ solutions
to $N!$ solutions, while in the Latin square problem the sampling
space is reduced from $L^{L\times L}$ to $(L!)^L$.
In the current work, the intermediate problems are associated with
different temperatures. An alternative way is to compute the density
of states $g(E)$ (i.e., the number of solutions with a particular
energy) by a random walk on the energy space \cite{b.muca, b.wl}.
However, we believe that the approach is less efficient than the one
used in this work. The reason is that while the energy range is
proportional to the system size $N$, it can be covered by much fewer
temperatures $\sim \sqrt N$. Thus it takes more time to estimate the
density of states than the partition function, especially for a
large system. Another advantage of the current method is that it
simplifies the design and application of efficient cluster
algorithms.  We expect the computational tool  to be broadly
applicable to other problems.

The authors acknowledge support from an NIH grant (R01-GM067801), an
NSF grant (MCB-0818353), and a Welch grant (Q-1512).


\end{document}